\documentclass[12pt]{iopart}

\usepackage{graphicx}
\usepackage[usenames]{color}
\newcommand{\beq}{\begin{equation}}
\newcommand{\eeq}{\end{equation}}
\newcommand{\beqa}{\begin{eqnarray}}
\newcommand{\eeqa}{\end{eqnarray}}
\newcommand{\ba}{\begin{array}}
\newcommand{\ea}{\end{array}}

\newcommand{\pr}{Phys. Rev.\ }

\begin{document}

\title{Study of a degenerate dipolar Fermi gas of $^{161}$Dy  atoms}

%\author{P. Muruganandam\footnote{anand@cnld.bdu.ac.in}} \affiliation
%%{Instituto de F\'{\i}sica
%Te\'orica, UNESP - Universidade Estadual Paulista, 01.140-070 S\~ao
%Paulo, S\~ao Paulo, Brazil} \affiliation{School of Physics,
%Bharathidasan University, Palkalaiperur Campus, Tiruchirappalli 620024,
%Tamilnadu, India}

\author{S. K. Adhikari %\footnote{adhikari@ift.unesp.br; URL  http://www.ift.unesp.br/users/adhikari  }
} 
\address{Instituto de F\'{\i}sica Te\'orica,
UNESP - Universidade Estadual Paulista, 01.140-070 S\~ao Paulo, S\~ao
Paulo, Brazil}

\begin{abstract} We  study properties of a single-component (spin polarized) degenerate dipolar 
Fermi gas of $^{161}$Dy atoms using a hydrodynamic description. 
Under axially-symmetric 
trapping we suggest reduced one- (1D) and two-dimensional (2D) description of the same
for  cigar and disk shapes, respectively. In addition to  a complete numerical solution of the 
hydrodynamic model we also consider a variational approximation of the same. For a trapped system 
under appropriate conditions, the 
variational approximation as well as the reduced   1D and 2D models are found to yield 
  results for shape, size  and chemical potential of the system in agreement with the full numerical 
solution of the three-dimensional (3D) model. For the uniform system we consider 
anisotropic sound propagation in 3D. An analytical result for anisotropic sound 
propagation in uniform dipolar degenerate Fermi gas is found to be in agreement with results of
numerical 
simulation in 3D.

 \end{abstract}


\maketitle

\section{Introduction}

After the experimental realization of Bose-Einstein condensate (BEC) of 
$^{52}$Cr \cite{pfau,4}, $^{164}$Dy \cite{7}, and $^{168}$Er \cite{8}
atoms with large magnetic dipolar interaction, there have been renewed interest in the study of  cold atoms, 
both theoretically and experimentally. The anisotropic long-range dipolar interaction acting in all partial waves in these atoms
is basically different from isotropic short-range  S-wave interaction acting in nondipolar atoms.  Polar bosonic molecules with much larger
permanent  electric
dipole moment  are also candidates  for possible BEC
experiments \cite{bosmol}.
Dipolar BECs have novel distinct properties. Due to the anisotropic  dipolar interaction, the stability of a
dipolar BEC depends on the scattering length as well as
the trap geometry \cite{4,12,14}.
The shock and sound waves also propagate anisotropically in a dipolar BEC
\cite{17,17A}. Anisotropic collapse has been observed and studied
in a dipolar BEC of $^{52}$Cr atoms \cite{18}.
Anisotropic rotons \cite{roton} and anisotropic critical superfluid 
velocity \cite{crit}
have been suggested and studied in dipolar BECs. Distinct stable checkerboard, stripe, and star configurations in dipolar BECs
have been identified in a two-dimensional (2D) optical lattice  as stable Mott insulator \cite{capo} as well as superfluid soliton \cite{star}
states. Anisotropic solitons in 2D have also been suggested in dipolar 
BECs \cite{anisol}. A new possibility of studying universal properties of dipolar BECs at unitarity has been suggested \cite{uni}.

After the realization of BEC of alkali-metal atoms, degenerate 
nondipolar gas of fermionic $^6$Li \cite{li}, $^{40}$K \cite{k}, and 
$^{87}$Sr \cite{sr} atoms were observed. Later, superfluid states of 
paired $^6$Li \cite{hulet} and $^{40}$K \cite{jin} Fermi atoms have been 
investigated. More recently a degenerate dipolar gas of fermionic 
$^{161}$Dy atoms with large magnetic dipole moment has been created and 
studied \cite{dipf}. Realization of quantum degeneracy in $^{161}$Dy 
atoms should be considered as a doorway for the study of anisotropic 
superfluidity in dipolar fermions. Fermionic polar molecules, such as 
$^{40}$K$^{87}$Rb, of large permanent electric dipole moment are also 
considered for this purpose \cite{fermol}. { The $^{40}$K$^{87}$Rb molecule in the 
singlet rovibrational ground state has an electric dipole moment of 0.6 Debye, 
thus leading to a dipolar interaction larger than in the case of $^{161}$Dy 
atoms by more than an order of magnitude \cite{pfau}.}
One advantage of studying the 
effect of dipolar interaction in a degenerate dipolar Fermi gas over 
that in a dipolar BEC is the remarkable stability of the degenerate 
Fermi gas. A BEC is usually fragile or short-lived for many experiments 
due to three-body loss through molecule formation. The three-body loss 
is highly suppressed in a degenerate Fermi gas due to Pauli repulsion 
among identical fermions. On the other hand a theoretical study of a BEC 
is much simpler than that of a degenerate Fermi gas \cite{frmp} due to 
the existence of a simple order parameter and a simple mean-field 
Gross-Pitaevskii equation in the former case.

A microscopic description of a degenerate Fermi gas is complicated due to the difficulties with the antisymmetrization of a many-fermion system and finding an appropriate simple order parameter. Drastic approximations are often necessary to achieve this goal for fermions. There  have been several different theoretical descriptions for a degenerate nondipolar Fermi gas \cite{dfer}.
There have also been a few studies of the degenerate dipolar Fermi gas
 employing different types of approximations \cite{fermidipth}.  However, for a description of some macroscopic observables of a degenerate Fermi gas, a simple hydrodynamic description \cite{hydro}
 often seems appropriate which does not require a precise antisymmetrization of the dynamics \cite{frmp,st,lca}. The effect of antisymmetrization is included approximately via an appropriate energy or Lagrangian density. A minimization of the energy leads to a well-founded variational approximation \cite{var}. An Euler-Lagrange equation derived from such a Lagrangian provides further improvement over the variational results. Here we present a description of the degenerate spin polarized Fermi gas based on such a hydrodynamic model \cite{hydro,st}.

The present theoretical formulation for the degenerate dipolar Fermi gas starts in section \ref{IIA}
with the standard equations of classical hydrodynamics \cite{hydro} with the appropriate equation of state including the kinetic energy of fermions filling the Fermi sea and the dipolar interaction among them.  The equivalent of the Thomas-Fermi (TF) approximation for the dipolar gas is obtained by setting the velocity field equal to zero in the hydrodynamic equations. An equivalent classical energy   density is written using the local-density approximation (LDA) \cite{frmp}.  A quantum pressure term is then introduced in the LDA energy density. With such a quantum pressure term, a nonlinear Schr\"odinger-type equation is derived for the density of fermions. For a moderate number of fermions (greater than 100 or so), the quantum pressure term is negligibly small and hence has insignificant effect on the result. {  However, the inclusion of the quantum 
pressure term allows one to write a dynamical equation for fermions in the form of a nonlinear Schr\"odinger 
equation 
to study the dynamics. The LDA or the TF 
approximation, on the other hand, allows only the investigation of static properties of the system. For example, these approximations cannot be used to study the sound propagation dynamics in fermions as in section 
\ref{IIIB}.}   
  In section \ref{IIB},  we present a Gaussian variational approximation for the problem described by the LDA energy density.
In section \ref{IIC},  simplified models are derived in reduced dimensions, appropriate for cigar and disk shapes of the degenerate dipolar gas when there is a strong trap in radial or in axial directions, respectively. In section \ref{IID}, using the hydrodynamic model, we obtain an analytic expression for the anisotropic sound velocity in the degenerate dipolar gas. 
In section \ref{IIIA}, we present numerical results for stationary properties $-$ shape, size, and chemical potential $-$ of a trapped degenerate Fermi gas of $^{161}$Dy atoms, and compare with results from appropriate models in reduced dimensions and variational approximation. In  section \ref{IIIB}, the anisotropic sound propagation in an infinite dipolar degenerate gas is demonstrated numerically  and the velocities so obtained are compared with the analytical results of section \ref{II} D. 
Finally, in section IV we present a 
summary of our study.

\section{Analytical Consideration}
\label{II}
\subsection{Hydrodynamic Model}
\label{IIA}

The normal one-component Fermi gas can be in the collisionless regime where collision is rare or in the collisional hydrodynamic 
regime where frequent collision due to the dipolar interaction  allow the system to settle to local equilibrium, where 
the system can be described by simple hydrodynamic equation rather than a detailed multi-particle description.    
Here we consider the system in such a configuration. A semi-quantitative estimate for the validity of a hydrodynamic description 
is given by the condition that relaxation time $\tau_R$ is small compared to the time scale $1/\omega$
defined by average trap frequency $\omega$  \cite{17,17A}.  For contact interaction this condition can be expressed 
in terms of the scattering length to measure the strength of atomic interaction \cite{17B}.  The 
strength of dipolar interaction is usually measured in terms of the convenient
length  $L_d\equiv 3a_{dd}= \mu_0\bar \mu^2 m/(4\pi \hbar^2)$, where $\bar \mu $ is the magnetic moment of an atom of mass $m$ and  $\mu_0$ the permeability of free space.
Using this length scale to measure dipolar interaction, the condition for the 
validity of hydrodynamic description can be expressed as \cite{17A,17B}
\begin{equation}\label{cond}
(\omega  \tau_R)^{-1}
\approx   \left( N^{1/3}\frac{L_d}{a_{\mathrm{ho}}}\right)^2F\left( \frac{T}{T_F}\right)  > 1,
\end{equation}
where $N$ is the number of atoms,
$a_{\mathrm{ho}}=\sqrt{\hbar/(m\omega)}$ the oscillator length, $T$  the temperature, $T_F$ the Fermi temperature and  
$F(x)$  an universal function and can be taken of the order of
unity \cite{17B} under experimental conditions. Hence for sufficently 
large  $N$ and/or sufficiently small   $a_{\mathrm{ho}}$ 
the system should enter the hydrodynamic regime. For $^{161}$Dy atoms of magnetic moment $10\mu_B$ with $\mu_B$ the Bohr magneton, the length $L_d\equiv 3 a_{dd}\approx 3\times 130 a_0$ where $a_0$ is the Bohr radius and for 
an oscillator length $a_{\mathrm{ho}}=0.25 $ $\mu$m and $N=5000$ the system is already in the hydrodynamic regime.  For polar Fermi molecule 
$^{40}$K$^{87}$Rb of electric dipole moment $d= 0.566$ Debye in the singlet 
rovibrational ground state \cite{fermol}, the length 
 $L_d =md^2/(4\pi\hbar^2\epsilon_0)\approx 3 \times 2000 a_0$ \cite{pfau}, where $\epsilon_0$ is the permitivity of free space, 
 and the system should enter 
the hydrodynamic regime with larger $a_{\mathrm{ho}} = 1$  $\mu$m (weaker trap)
and much smaller $N $ of few hundreds.

At sufficiently low temperature the    normal Fermi gas enters  
the degenerate phase.
Most macroscopic properties, like shape, density, chemical potential, sound propagation etc., of this system can be described 
by the  Landau hydrodynamical equations \cite{hydro}.
A velocity field 
is introduced as the gradient of a velocity potential $\Phi$ of flow, usually related to phase of the order parameter,  by 
$
{\bf v}({\bf r},t)=   \nabla \Phi({\bf r},t),
$
subject to the irrotational condition 
$\nabla \times {\bf v}({\bf r},t)=0$.
% where $\tilde m$ is the effective mass 
%responsible for flow and $\Phi$ is the velocity potential of flow. To be consistent with the correct quantum pressure term, which we introduce later, and with Galilei invariance and correct sound velocity, we shall choose 
%$\tilde m =2m$. 
The continuity  and the flow equations are then given by
\cite{hydro}
\begin{eqnarray}\label{h1}
&&\frac{\partial n({\bf r},t)}{\partial t}+\nabla \cdot [n({\bf r},t) \bf v({\bf r},t) ]=0,\\
&& m\frac{\partial \bf v(\bf r,t)}{\partial t}+\nabla\left[ 
\frac{ m {{\bf v(r},t)}^2}{2}+\mu(n,{\bf r})+
V({\bf r})\right]=0,\label{den}
\end{eqnarray}
where $n({\bf r},t)$ is density at space point $\bf r$ and time $t$,
$V(\bf r)$ is an external trap, usually taken %in the axially-symmetric form 
as 
\begin{equation} \label{pot}
V({\bf r})=\frac{1}{2}m\omega^2 (\nu^2 \rho^2+\lambda^2 z^2),
\end{equation} 
with $\omega$ the trap frequency and $\nu$ and $\lambda$ are anisotropy parameters.
%For a uniform system $V({\bf r})=0$. 
The bulk chemical potential
$\mu(n, \bf r)$ is   determined by the equation of state of the 
uniform Fermi gas of dipolar atoms and is given by 
\begin{eqnarray}\label{chemx}
\mu(n,{\bf r})=  
\frac{\hbar^2 [6\pi^2 n({\bf r})]^{2/3} }{2m} +
\int d{\bf r'} n({\bf r'}) U_{dd}({\bf r-r'}),
\end{eqnarray}
where the first term on the right-hand side (rhs) is the Fermi energy $E_F\equiv{\hbar^2 
(6\pi^2 n)^{2/3} }/({2m}) 
$ \cite{frmp}
and the last 
term is the contribution from dipolar interaction energy \cite{st}, with $U_{dd}(\bf r -r ')$ 
is the dipolar potential. 

%The hydrodynamical equations (\ref{h1}) and (\ref{den}) are not appropriate to study the properties of the system requiring microscopic treatment. 
The hydrodynamic description is
valid for a macroscopic
observable with its  characteristic excitation wave-length $\lambda$  
much larger than the healing length \cite{frmp}. A safe condition to satisfy this criterion is to take the wave length to be much larger than  
de Broglie wavelength at the
Fermi surface, i.e., \cite{frmp,am}
\begin{equation}\label{con}
\lambda  \gg 2\pi/k_F 
\end{equation}
with Fermi momentum $k_F$ defined by 
$E_F=\hbar^2k_F^2/2m$. 
{ Truly speaking, a degenerate Fermi
gas may not be fully irrotational and allows rotational components
in the velocity field, not allowed in the present 
hydrodynamic formulation. This fact should influence only the rotational
properties \cite{frmp} of the degenerate Fermi gas not
considered in this paper. We also assume the absence of any velocity dependent frictional force.}
%This will be satisfied by the shapes and sizes of the
%trapped degenerate Fermi gas as considered here. 
%This condition is similar in the hydrodynamic approximation in classical
%rarefied gases: the scale of the flow should be
%much larger than the mean-free path.
%{It is now realized \cite{frmp} that 
%the macroscopic properties of a 
%noninteracting degenerate Fermi gas is also governed by the same %hydrodynamical equations, although such a degenerate Fermi gas in not fully %irrotational and allows 
%rotational components in the velocity field, not allowed in superfluid hydrodynamics. This fact should influence only the rotational properties \cite{frmp}
%of the degenerate Fermi gas not considered in this paper.  }

An approximate TF profile for  density  can now be obtained by setting  velocity   ${\bf v}=0$ in   (\ref{den}), when \cite{frmp}
\begin{equation}\label{tf}
\mu(n,{\bf r})+V({\bf r})=\mu_0^{\mathrm{TF}},
\end{equation}
where $\mu_0^{\mathrm{TF}}$ is the chemical potential of the trapped gas.  When   (\ref{tf}) is solved 
subject to the appropriate normalization condition, we obtain both the chemical potential $\mu_0^{\mathrm{TF}}$ and  the density $n(\bf r)$.
% The TF profile  goes to zero abruptly with discontinuous 
%derivative beyond a spatial extension. 
%Nevertheless, the TF result has been useful for providing a reasonably accurate 
%density, specially for systems with a large number of atoms. 

There is an equivalent description of the trapped degenerate Fermi gas in the  LDA, based on the assumption that, locally the dipolar Fermi gas would behave like a uniform gas, so that the energy density can be written as the energy of the uniform system times the local density \cite{frmp}.  
For the degenerate dipolar Fermi system,  the classical energy 
density per particle  is given by \cite{fermidipth,st}
\begin{eqnarray}
 \label{lagcl} {\cal E}_{\mathrm{cl}}({\bf r})&&= \frac{3}{5} 
\frac{\hbar^2}{2m}(6\pi^2N)^{2/3} [n({\bf r})]^{5/3}+V({\bf r}) n({\bf r})\nonumber \\ 
&&+N\frac{1}{2}\int d{\bf r'}n({\bf r}) n({\bf r'}) U_{dd}({\bf r-r'}), \end{eqnarray} 
where $n(\bf r)$ is the density of fermions normalized as $\int d{\bf r}n({\bf 
r}) =1$.    
The first term on the rhs of   (\ref{lagcl}) is the total 
kinetic energy of the spin polarized fermions filling all levels up to the Fermi sea, the second 
term is energy in the trap, and the third term describes the dipolar interaction. A minimization of the classical energy (\ref{lagcl})
leads to the TF condition (\ref{tf}).

As the degenerate Fermi gas is a quantum system, a quantum pressure term  when included in   (\ref{lagcl})
yields the following expression  for net energy density 
\begin{eqnarray}\label{grad}
\label{lag} &&{\cal E}({\bf 
r})= \frac{\hbar^2}{8m}|\nabla_{\bf r}\sqrt{n(\bf r)}|^2 +{\cal E}_{\mathrm{cl}}({\bf r}).
 \end{eqnarray}
This gradient correction term \cite{grad} to the TF energy density (\ref{lagcl})
takes into account the additional kinetic 
energy due to spatial variation of density (near the surface).
% but contributes little to the energy of the degenerate Fermi gas compared to the 
%dominant volume term $-$ the first term on the rhs    of   (\ref{lagcl}).
Such a surface term was first considered by von Weisz\"acker \cite{von,toigo,kohn} in the description  a large nuclei. 
%With a
% similar quantum pressure term  in the superfluid bosonic 
%energy density, one can establish a complete equivalence between the 
%well-established mean-field quantum mechanical 
%GP equation and a quantum superfluid hydrodynamical model \cite{rmp}.  
%The existence of this  gradient correction term in the energy density 
%of a quantum system is essential for a correct description \cite{grad}. 
% Although the gradient form 
%of this term is consistent with quantum mechanics, the coefficient in this term  is decided
%phenomenologically. 
Previous descriptions of a degenerate Fermi gas considered different 
coefficients in this term
\cite{dfer,grad}.  
%In this work we take the coefficient
%$\hbar^2/8m$, which is consistent with a 
%Galilei-invariant 
%hydrodynamic description of a Fermi superfluid governed by the 
%mass $2m$ of a 
%Cooper pair \cite{as}, 
The energy density (\ref{lag}) has 
successfully used in many problems of Fermi gas 
\cite{toigo,kohn,as,many}. 
%The quantum pressure term
%in the energy density  for the Cooper pair 
%of mass $2m$  generates the $\hbar^2/8m$ factor in the energy density of 
%fermions, viz.,   (35) of Ref. \cite{as}.
%The quantum pressure term only contributes for small number of fermions; %for 
%a large number of fermions this term is negligibly small. 
%For a small number of 
%fermions this term with coefficient $\hbar^2/8m$ has yielded \cite{as}
%results for the energy of a small trapped Fermi system 
%in close agreement with the accurate Monte Carlo calculation \cite{mc} as %well as in approximate agreement 
%with LDA. 
%The use of this factor has also led to a satisfactory description 
%of density of a trapped Fermi gas \cite{adhik} and of the dynamics of collision of 
%two Fermi clouds \cite{toigo}. This factor also leads to a better 
%representation of a Kohn-Sham calculation for electrons \cite{kohn}.
%This is why we shall adopt the quantum pressure term with the 
%coefficient $\hbar^2/8m$ in this study. 

%An advantage of using the quantum pressure term in   (\ref{lag})
%is that 
%an equivalent Lagrangian density ${\cal L}({\bf r})={\cal E}({\bf r})-\mu_0n(\bf r)$,
%can be defined with the quantum pressure term included, which 
%yields a nonlinear Schr\"odinger equation 
%for density $n(\bf r)$ in close analogy with the mean-field GP
%equation for bosons.  

With the  Lagrangian  density ${\cal L}({\bf r})={\cal E}({\bf r})-\mu_0n(\bf r)$ % of     (\ref{lag}) 
the Euler-Lagrange equation   %for the three-dimensional (3D) system 
is given by  \cite{as}
\begin{eqnarray}  \label{gp3d} 
 \mu_0 \sqrt{n{(\bf r)}}
&& =  \biggr[ -\frac{\hbar^2\nabla^2 }{8m} + V({\bf r})
+
\frac{\hbar^2}{2m}[6\pi^2 N n({\bf r})]^{2/3}\nonumber \\
&&+  N\int U_{dd}({\bf r -r'})n({\bf r'})d{\bf r'}
\biggr] \sqrt{n({\bf r})}, 
\end{eqnarray} 
with $\mu_0$ the chemical potential.
%The quantum pressure term has led to the $\nabla^2$ term in the nonlinear 
%Schr\"odinger equation (\ref{gp3d}). 
The derivarive term in   (\ref{gp3d}) term contributes much less than the 
dominant ``Fermi energy" term $\hbar^2[6\pi^2Nn({\bf r})]^{2/3}/(2m)$ and its neglect 
 leads to the  TF relation (\ref{tf}). The   dipolar interaction in   (\ref{gp3d}) is taken as 
$
 U_{dd}({\bf R}) = 3
a_{dd}\hbar^2(1-3\cos^2	\theta)$ $/(mR^3),$ $\quad {\bf R=r-r'}
$ where $	\theta$ is the angle between the $\bf R$ and the polarization 
direction $z.$

The condition (\ref{cond}) refers to the validity of a 
hydrodynamic  description of the system ($\bf v \ne 0$) \cite{17A,17B}. A reliable stationary description obtained 
using   (\ref{gp3d}) for $\bf v =0$,
of density and other macroscopic properties, such as sound 
velocity, as considered in this paper, 
can be obtained for a smaller number of fermions
consistent with condition (\ref{con}) \cite{frmp,am}, provided the contribution of the kinetic energy term $\hbar^2[6\pi^2N n({\bf r})]^{2/3}
/(2m)$ in   (\ref{gp3d}) is much larger than that of  other terms. 
 
%The constant $a_{dd}
%=\mu_0\bar \mu^2 m /(12\pi \hbar^2)$ 
%is a length characterizing the strength of dipolar interaction with
%$\bar \mu$ the magnetic dipole moment of a single atom, and $\mu_0$ 
%the permeability of free space.
%The experimental
%value of $a_{dd}$
%for 
%bosonic $^{52}$Cr ($\bar \mu=6\mu_B$) \cite{4},  $^{168}$Er 
%($\bar\mu=7\mu_B$) \cite{7}, and $^{164}$Dy   ($\bar \mu=10\mu_B$)
%\cite{8}
%atoms 
%  are 
%approximately taken as 
%$15a_0$ \cite{4}, 
%$65a_0$, and $130a_0$, respectively,
%with  $a_0$ the Bohr 
%radius and  $\mu_B$ a Bohr magneton.  
% For 
%fermionic $^{161}$Dy atoms, 
%with $\bar \mu=10\mu_B,$ is
%$a_{dd}\approx 130a_0$ \cite{dipf}, with $a_0$ the Bohr 
%radius and  $\mu_B$ a Bohr magneton. 

It is convenient to write a dimensionless form of   (\ref{gp3d})
with the potential (\ref{pot})
as 
\begin{eqnarray}  \label{3d1} 
 \mu_0 \sqrt{n{(\bf r)}}
&& =  \biggr[ -\frac{\nabla^2 }{8} + \frac{1}{2}(\nu^2\rho^2+\lambda^2z^2)
+
\frac{1}{2}[6\pi^2 N n({\bf r})]^{2/3}\nonumber \\
&&+  3a_{dd}N\int 
%U_{dd}({\bf r -r'})
\frac{1-3\cos^2 	\theta }{R^3}
n({\bf r'})d{\bf r'}
\biggr] \sqrt{n({\bf r})}, 
\end{eqnarray} 
where energy, length and density are expressed in units of 
$\hbar\omega$, $l_0=\sqrt{\hbar/m\omega}$ and $l_0^{-3}$.

\subsection{Variational Approximation}
\label{IIB}

  The energy density corresponding to the dimensionless equation (\ref{3d1})
 can be written as 
\begin{eqnarray}
{\cal E}({\bf r})&&= \frac{1}{8}|\nabla_{\bf r}\sqrt{n(\bf r)}|^2 
+\frac{3a_{dd}N}{2}\int d{\bf r'}n({\bf r})
n({\bf r'}) \frac{1-3\cos^2 	\theta }{R^3}
%U_{dd}({\bf r-r'})
\nonumber \\
&&+\frac{1}{2}(\nu^2\rho^2+\lambda^2 z^2)n({\bf r})+\frac{3}{10}
(6\pi^2N)^{2/3}
[n({\bf r})]^{5/3}
.
\end{eqnarray}
%This is just the energy density (\ref{lag}) in dimensionless units. 
A variational approximation for the problem  can be obtained with the following Gaussian ansatz for density \cite{var}
\begin{eqnarray}
n({\bf r})= \frac{1}{\pi^{3/2} w_\rho^2 w_z}  \exp\left[-\frac{\rho^2}{w_\rho^2} -\frac{z^2}{w_z^2}  \right],
\end{eqnarray}
where $w_\rho$ and $w_z$ are the variational widths along radial $\rho$ and axial $z$ 
directions. %With this density the root-mean-square (rms) sizes $\langle x \rangle$
%and $\langle z \rangle$ are, respectively, $w_\rho/\sqrt 2$ and $w_z/\sqrt 2$.  
With this density,
the effective energy per particle  of the system 
$E=\int d{\bf r}{\cal E}({\bf r})$ 
is 
\begin{eqnarray}\label{energy}
E&&= \frac{1}{8}\left[\frac{1}{w_{\rho}^2}+
\frac{1}{2w_{z}^2}
\right]
+\left[\frac{\nu ^2 w_{\rho}^2}{2}+\frac{\lambda^2 w_z^2}{4}\right]
\nonumber \\
&&-\frac{Na_{dd}f(\kappa)   }{\sqrt{2\pi} w_{\rho}^2w_{ z}} 
+ \sqrt{\frac{3}{5}}  \frac{9 (6\pi^2  N)^{2/3} }{50
\pi w_{\rho}^{4/3} w_{z}^{2/3}},
\end{eqnarray}
where $ \kappa = w_\rho/w_z$ and 
\begin{eqnarray}
f(\kappa)=\frac{1+2\kappa^2}{1-\kappa^2} -\frac{3\kappa^2\mbox{tanh}^{-1}
\sqrt{1-\kappa^2}}{(1-\kappa^2)^{{3/2}}}. \label{eqn:fkappa}
\end{eqnarray}
The  variational equations are obtained by minimizing the energy 
(\ref{energy}) by $\partial E/\partial w_z= \partial E/\partial 
w_\rho=0$ \cite{4,var}:
\begin{eqnarray}\label{var1}
 {w_{\rho}}\nu^2  &&=
\frac{1}{4w_{\rho}^3} -\frac{
a_{dd}}{\sqrt{2\pi}} \frac{Ng(\kappa)}{w_{\rho}^3w_{z}}
+\sqrt{\frac{3}{5}}\frac{6(6 \pi^2 N)^{2/3}}
{25\pi w_{z}^{2/3}w_{\rho}^{7/3}}
, \\  {w}_{z} \lambda ^2 &&=
\frac{1}{4w_{z}^3}- \frac{ a_{dd}}{\sqrt{2\pi}}
\frac{2Nh(\kappa) }{w_{\rho}^2w_{z}^2} 
+\sqrt{\frac{3}{5}}\frac{6(6 \pi^2 N)^{2/3}}{25\pi w_{z}^{5/3}
w_{\rho}^{4/3}},\label{var2}
\end{eqnarray}
where 
%\begin{subequations}
\begin{eqnarray}
&& g(\kappa) = \frac{2-7\kappa^2-4\kappa^4}{(1-\kappa^2)^2} +
\frac{9\kappa^4\mbox{tanh}^{-1}\sqrt{1-\kappa^2}}{(1-\kappa^2)^{5/2}}, \\
&& h(\kappa) = \frac{1+10\kappa^2-2\kappa^4}{(1-\kappa^2)^2} -
\frac{9\kappa^2\mbox{tanh}^{-1}\sqrt{1-\kappa^2}}{(1-\kappa^2)^{5/2}}.
\end{eqnarray}
%\end{subequations}
%The contribution of the quantum pressure term in   (\ref{var1}) and (\ref{var2}) is in the first terms on the rhs $1/(4w_\rho^3)$ and $1/(4w_z^3)$. The contribution of the fermion kinetic energy term is contained in the last term of these equations.  For a moderate value of $N$, the coefficient of the last terms is much larger than that of the quantum pressure term, as 
%\begin{equation}
%\sqrt{\frac{3}{5}}\frac{6(6 \pi^2 N)^{2/3}}
%{25\pi } \gg \frac{1}{4},
%\end{equation}
%demonstrating the smallness of the quantum pressure term.
%The stationary widths $w_\rho$ and $w_z$ are determined by solving   (\ref{var1})
%and (\ref{var2}). 
The chemical potential $\mu_0$ of the system per particle is
\begin{eqnarray} \label{muvar}
\mu_0 &&= \frac{1}{8}\left[\frac{1}{w_{\rho}^2}+
\frac{1}{2w_{z}^2}
\right]
+\left[\frac{\nu^2 w_{\rho}^2}{2}+\frac{\lambda^2 w_z^2}{4}\right]
\nonumber \\
&&-\frac{2Na_{dd}f(\kappa)   }{\sqrt{2\pi} w_{\rho}^2w_{ z}} 
+ \sqrt{\frac{5}{3}}  \frac{9 (6\pi^2  N)^{2/3} }{50
\pi w_{\rho}^{4/3} w_{z}^{2/3}}.
\end{eqnarray}
%can be compared with the similar expression for energy (\ref{energy}).

\subsection{Approximate density for cigar and disk shapes}
\label{IIC}

In many situations of experimental interest, the   Fermi gas
could be subject to %an axially-symmetric trap, could be such that there is 
a strong trap either in the polarization $z$ direction or in the 
transverse radial $\rho$ plane.
% perpendicular to the polarization direction. 
The system then has a one-dimensional (1D) cigar  or 2D disk shape, respectively. In such cases simplified equations in lower dimensions could be very useful  \cite{luca}.

First we consider the reduced 1D equation for a cigar shape.  
%To reduce   (\ref{3d1}) to a 1D form 
We assume that there is a strong trap 
in the $x-y$ plane %perpendicular to the polarization direction $z$
and that the 
density in this  plane is given by the Gaussian ground state 
\cite{luca}
$n(\rho)=\exp(-\rho^2/d_\rho^2)/(\sqrt \pi d_\rho),  d_\rho = \sqrt {1/(2\nu)}$
of the trap  $\nu^2\rho^2/2$,
so that the 3D density $n({\bf r})$ satisfies
\begin{eqnarray}
\sqrt {n({\bf r})}\equiv  \sqrt{n(\rho)}\sqrt{n(z)}= \frac{ \sqrt{n(z)}
}{\sqrt \pi d_\rho}\exp\left[-
\frac{\rho^2}{2d_{\rho}^2}
\right] .
\end{eqnarray}
Substituting this density in   (\ref{3d1}),
and multiplying this equation by the corresponding Gaussian $\sqrt{n(\rho)}$
and integrating out the $\rho$ dependence
we obtain the reduced 1D equation \cite{1D,lp}     
\begin{eqnarray}  \label{1D} 
&&\mu_{1D} \sqrt{n(z)}
 =  \biggr[ -\frac{\partial_z^2}{8} +\frac{\lambda^2 z^2}{2}
+   \frac{3[6  N n(z)]^{2/3}}{10 d_\rho^{4/3}  }
 \nonumber \\
&&+  \frac{2a_{dd}N}{d_\rho^2}\int_{-\infty}^{\infty}
\frac{dk_z}{2\pi}e^{ik_z z}n(k_z)s_{1D}\left( \frac{k_zd_\rho}{\sqrt 2} \right)  \biggr]\sqrt{n(z)}
,\\
\label{1Dvdd}
&&s_{1D}(\zeta)=\int_0^\infty du \frac{2\zeta^2 -u}{u+\zeta^2} 
   e^{-u},  
\end{eqnarray}
where 
$n(k_z)=\int_{-\infty}^{\infty} e^{-ik_z z}n(z) dz,$
and  $\mu_{1D}$ is the chemical potential.
% Equation (\ref{1D}) is the reduced 1D equation appropriate for a cigar-shaped
%dilute dipolar degenerate Fermi gas.  
An approximate variational solution of   (\ref{1D}) is possible with the following ansatz for density \cite{lp}
%\begin{equation}
 $ n(z)={1}/({\sqrt \pi  w_z})\exp[ -{z^2}/{ w_z^2}  ],$
%\end{equation}
while the width $w_z$ is determined by solving   (\ref{var2}) with $w_\rho =d_\rho$ and $\kappa= d_\rho/w_z$. 

Next we consider the reduced 2D equation suitable for a disk shape 
 with a strong axial trap. The dipolar Fermi gas is assumed 
to be in the ground state \cite{luca} $n(z)=\exp(-z^2/d_z^2)/(\sqrt \pi d_z), d_z=\sqrt{1/(2\lambda)}$,
of the axial trap $\lambda^2z^2/2$
and the 3D density can  be approximated as 
\begin{eqnarray}
\sqrt {n({\bf r})}\equiv  \sqrt{n(\rho)}\sqrt{n(z)}= \frac{\sqrt{n(\rho)}
}{ \pi^{1/4} \sqrt{d_z}}\exp\left[-
\frac{z^2}{2d_z^2}
\right] .
\end{eqnarray}
Using this ansatz in   (\ref{3d1}),
and multiplying   by the corresponding Gaussian $\sqrt{n(z)}$
and integrating out the $z$ dependence 
we get   
 the   effective 2D equation \cite{lp,2D}
\begin{eqnarray}
&&\mu_{2D}\sqrt{n(\rho)}=
\biggr[ -\frac{\nabla^2_\rho}{8}+\frac{1}{2}\nu^2 \rho^2+
\sqrt{ \frac{3}{5}  } \frac{[6N 
n(\rho)]^{2/3}\pi}{2d_z^{2/3}} \nonumber
\\
\label{2D}
&&   + \frac{4\pi a_{dd}N}{\sqrt{2\pi}d_z}
\int\frac{d^2k_\rho}{(2\pi)^2}e^{i{\bf k_\rho . 
\rho}}
n({\bf k_{\rho}})  h_{2D}\left( \frac{k_\rho d_z}{\sqrt 2}  \right)
\biggr]  \sqrt{n(\rho)},
\end{eqnarray}
where 
$n({\bf k_{\rho}})=\int e^{i\bf k_\rho . \rho }n(\rho) d^2\rho,
h_{2D}(\xi)=2-3\sqrt \pi \xi e^{\xi^2} [1-\mbox{erf}(\xi)].$
% Equation (\ref{2D}) is the reduced 2D equation appropriate for a disk-shaped
%dilute dipolar degenerate Fermi gas.    
 An approximate variational solution of   (\ref{2D}) is possible with the following ansatz for density \cite{lp}
$
  n(\rho)={1}/{( \pi  w_\rho^2)}\exp[ -{\rho^2}{ w_\rho^2}  ],
$
while the width $w_\rho$ is determined by solving   (\ref{var1}) with $w_z =d_z$ and $\kappa= w_\rho/d_z$.

\subsection{Sound propagation in a uniform  Fermi gas}
\label{IID}

To find the sound velocity in a uniform Fermi gas, we evaluate the
Bogoliubov spectrum using the linearized hydrodynamic equations. We consider 
a {\it stationary} Fermi gas in a box with periodic boundary condition with the trap $V$ fixing just the allowed plane wave solution. Then   (\ref{den}), after the inclusion of the gradient term of   (\ref{grad}),
reduces to
\begin{equation}\label{h3}
-\frac{\hbar^2\nabla^2 \sqrt{n({\bf r},t)}}{8m\sqrt{n({\bf r},t)}}+\mu(n,{\bf r})+
m \frac{\partial \Phi({\bf r},t)}{\partial t}=0.
\end{equation}
Now we allow small perturbation in $n$ and $\Phi$ around their equilibrium 
values $n_0$ and $\Phi_0$ by $n({\bf r},t) \approx n_0+\bar n({\bf r},t)$ 
and $\Phi  ({\bf r},t)\approx \Phi_0+\bar \Phi({\bf r},t)$, then   (\ref{h1}) leads to
\begin{equation}\label{h4}
\frac{\partial \bar n({\bf r},t)}{\partial t}+n_0\nabla^2 \bar \Phi({\bf r},t) =0.
\end{equation}
In   (\ref{h3}), we need to use $\sqrt {n({\bf r},t)}  \approx \sqrt{n_0}+ 
\bar n({\bf r},t)/(2\sqrt{n_0})$ and $ [n({\bf r},t)]^{2/3}  \approx n_0^{2/3}+ 2\bar n({\bf r},t)/(3{n_0}^{1/3})$, while $\mu(n,{\bf r})\approx 
\tilde \mu_0+\bar\mu (n,{\bf r})$ with  
\begin{equation}\label{h5}
\bar \mu(n,{\bf r})=\frac{\hbar^2(6\pi^2)^{2/3}\bar n}{3mn_0^{1/3}}+\int
d{\bf r'}\bar n({\bf r'},t)U_{dd}({\bf r-r'}),
\end{equation} 
where $\tilde  \mu_0$ is the stationary value of $\mu$. Then   (\ref{h3}) leads to
\begin{equation}\label{h6}
-\frac{\hbar^2\nabla^2 {\bar n({\bf r},t)}}{16mn_0}+\bar \mu(n,{\bf r})+
m \frac{\partial\bar  \Phi({\bf r},t)}{\partial t}=0.
\end{equation}
Assuming the perturbations $\bar n$ and $\bar \Phi$ in plane-wave forms 
$\bar n =\bar n_0\exp[i({\bf k \cdot r})-\omega t]$
and $\bar \Phi =\bar \Phi_0\exp[i({\bf k \cdot r})-\omega t],$ $\bar \mu$ has the form $\bar\mu =\bar n_0 \bar \mu_0\exp[i({\bf k \cdot r})-\omega t]$ with 
\begin{equation}\label{h10}
\bar \mu_0= \frac{\hbar^2(6\pi^2)^{2/3}}{3mn_0^{1/3}}+   \frac{4\pi  a_{dd}\hbar^2}{m} (3\cos^2\theta -1),  
\end{equation}
where $\theta$ is the  angle  between  the propagation direction and  polarization direction 
$z$, and the last term in   (\ref{h10}) is just the Fourier transform of the dipolar potential. Then  
   (\ref{h4}) and (\ref{h6}) become:
\begin{eqnarray}
&&-i\omega \bar n_0  -n_0k^2\bar \Phi_0 =0,\\
&&\left[ \frac{\hbar^2k^2}{16mn_0} + \bar\mu_0  \right]\bar n_0 -i\omega m \bar\Phi_0 =0. 
\end{eqnarray}
The condition of existence of the solution to this set of equations leads to 
the Bogoliubov spectrum 
\begin{equation}
\epsilon_k\equiv \hbar \omega = \sqrt{\frac{\hbar^2k^2}{4m}\left[
\frac{\hbar^2k^2}{4m}+4n_0\bar \mu_0  \right]
}.
\end{equation}
The sound velocity, defined as $c_s(\theta)=\lim_{k\to 0}(\epsilon_k/\hbar k)$ 
for a uniform dipolar Fermi gas can be written as 
\begin{equation}\label{vel}
c_s(\theta) = \sqrt{\frac{v_F^2}{3}+ \frac{4\pi n_0 a_{dd}\hbar^2}{m^2}(3\cos^2\theta-1)},
\end{equation}
where  $v_F\equiv \hbar k_F/m =\hbar(6 \pi ^2n_0)^{1/3}/m$ is the 
Fermi velocity. 
 For a nondipolar Fermi gas ($a_{dd}=0$),   (\ref{vel}) leads to the well-known 
 velocity of $v_F/\sqrt 3$ \cite{frmp,lca}.

{The angle-dependent second term under the square root in  (\ref{vel}) is responsible 
for anisotropic sound velocity. Specifically, for $\theta > 54.73^\circ$ degrees, this term is negative 
corresponding to a decrease in sound velocity. For large dipolar interaction, for $\theta$ greater than 
a critical value and for large density $n_0$
the sound velocity could be imaginary corresponding to no propagation. However, in this study we shall only consider moderate values of density and dipolar interaction, that allow anisotropic sound propagation in all directions.  
}
\section{Numerical Results}

\label{III}
\subsection{Stationary properties of trapped  dipolar Fermi gas}

\label{IIIA}

For a trapped 3D Fermi system we solve   (\ref{3d1}) numerically after discretization \cite{CPC}. The divergence of the dipolar term at short distances 
has been handled by treating this term in momentum ($\bf k$) space \cite{12}. 
%The dipolar integral 
 %in momentum space is tackled by the following convolution integral 
%\cite{12}
%\begin{eqnarray}&
%\int d{\bf r'}U_{dd}({\bf r-r'})n({\bf r'})
%=\int \frac{d{\bf k}}{(2\pi)^ 3}\exp(-i{\bf k \cdot r})
% U_{dd}({\bf k})n({\bf k}). \nonumber
%\end{eqnarray}
%The Fourier transformation (FT) is defined by
%\begin{eqnarray}
%&
%A({\bf k})=\int d{\bf r} B({\bf r})\exp(i {\bf k \cdot r}),
%\\&
% B({\bf r})=\frac{1}{(2\pi )^3}
%\int d{\bf k} A({\bf k})\exp(-i {\bf k \cdot r}).
%\end{eqnarray}
% The FT $U_{dd}({\bf k})$ of the 
%dipolar potential is known analytically \cite{12,pfau}:
%\begin{eqnarray}
%U_{dd}({\bf k})=4\pi a_{dd}\left(\frac{3k_z^2}{{\bf k}^2} -1   \right).
%\end{eqnarray}
% The FT $n({\bf k})$ of density 
%is calculated numerically by a fast FT (FFT) routine.  The inverse FT is 
%also evaluated numerically by a FFT routine. The whole procedure is 
%performed in 3D Cartesian coordinate system irrespective of the underlying 
%trap symmetry. 
For numerical calculation in section \ref{IIIA}, we consider a degenerate Fermi gas of  $^{161}$Dy atoms with $a_{dd} \approx 130a_0$ and employ
the oscillator length $l_0 = 1$ $\mu$m.

\begin{figure}%[!ht]
\begin{center}
\includegraphics[width=.7\linewidth]{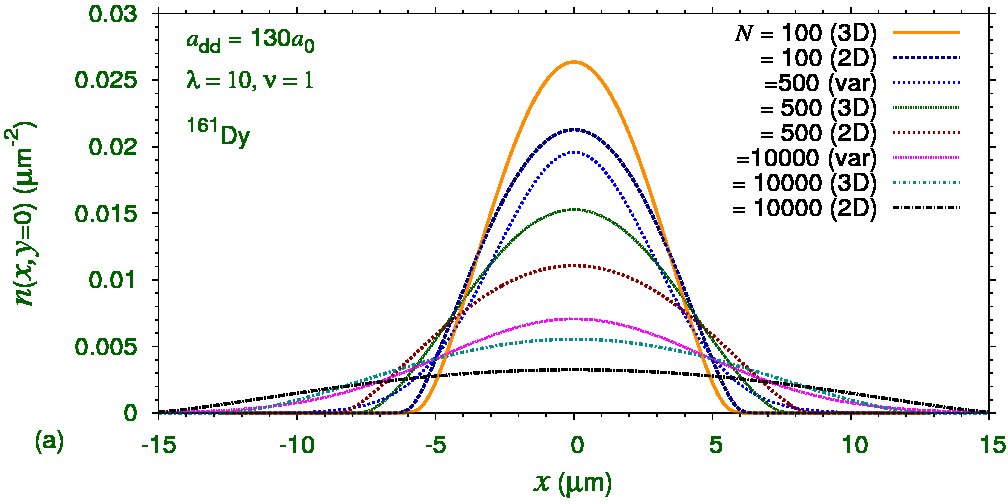}
\includegraphics[width=.7\linewidth]{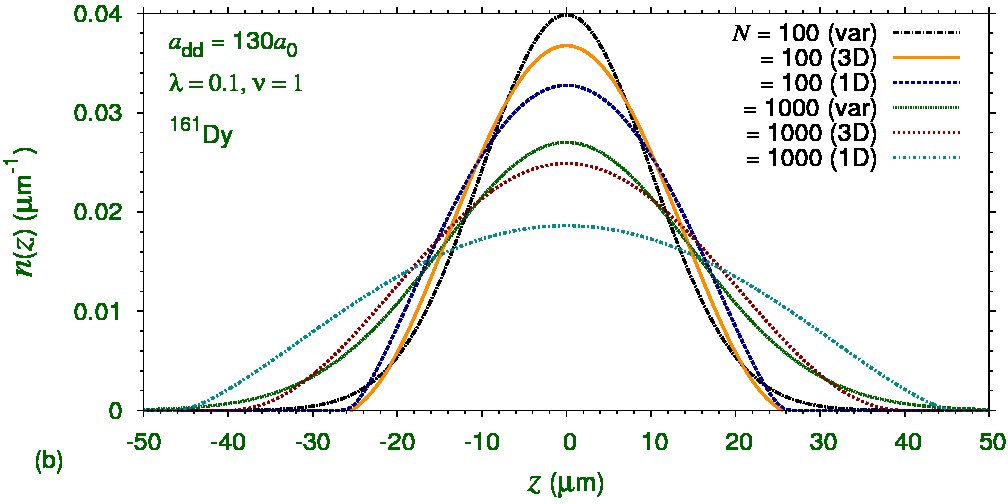}
\end{center}

\caption{  (a)  Radial density of a disk-shaped 
Fermi gas of $^{161}$Dy atoms
in $x-y$ plane 
$n(x,y=0)=\int_{-\infty}^{\infty} dz n(x,y=0,z)$ from the 3D   
(\ref{3d1}) for  $\nu=1, \lambda=10$ 
compared 
with its variational (var) approximation and
the solution of the reduced 2D   (\ref{2D}) for   
$N=100, 500, 10000$. 
(b) Axial density of a cigar-shaped 
Fermi gas of $^{161}$Dy atoms
along polarization $z$ axis 
$n(z)=\int_{-\infty}^{\infty}   \int_{-\infty}^{\infty} dx dy n(x,y,z)$ from the 3D   
(\ref{3d1}) for  $\nu=1, \lambda=0.1$ 
compared 
with its variational (var) approximation and
the solution of the reduced 1D   (\ref{1D}) for  
$N=100, 1000$. 
}

\label{fig1}
\end{figure}

\begin{figure}[!ht]
\begin{center}
\includegraphics[width=.7\linewidth]{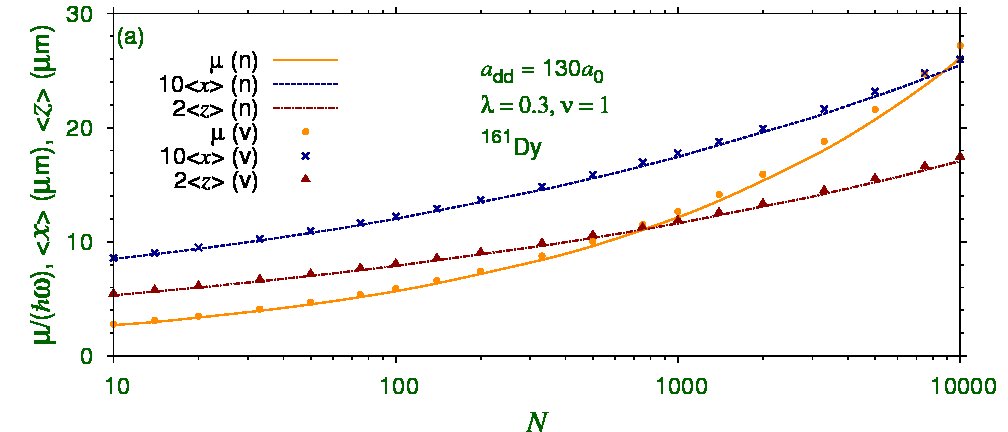}
\includegraphics[width=.7\linewidth]{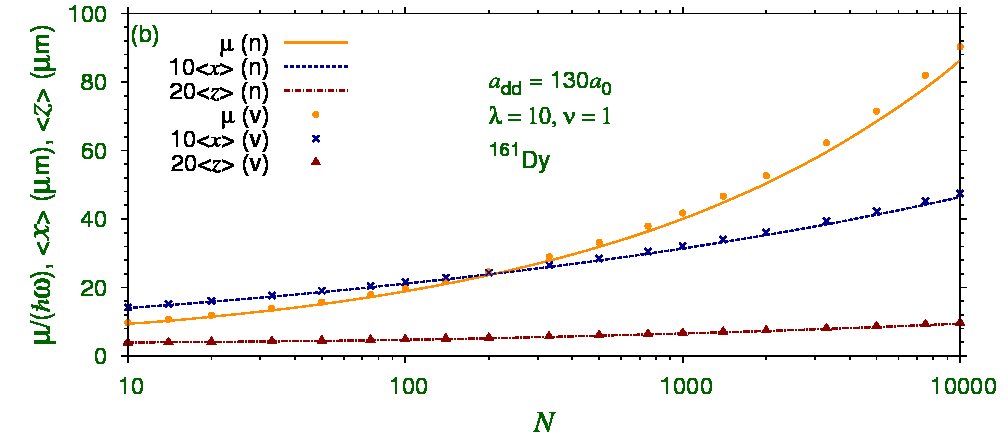}
\includegraphics[width=.7\linewidth]{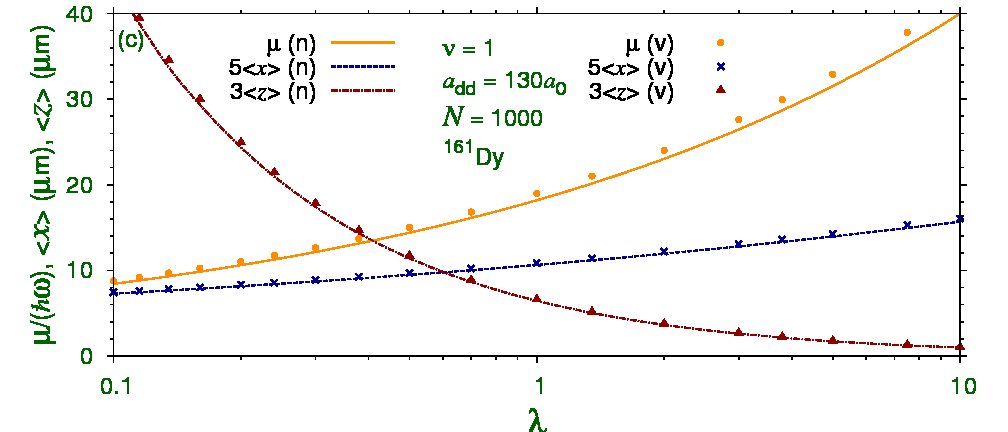}
\end{center}

\caption{  (a)  Reduced chemical potential $\mu/(\hbar \omega)$, 
and rms sizes $\langle x \rangle $, and $\langle z \rangle $ of a cigar-shaped Fermi 
  gas of $^{161}$Dy atoms with 
$\nu=1,\lambda =0.3$ versus   $N$ from a numerical (n) solution of the 
3D   (\ref{3d1}) and variational  (v) approximation [   (\ref{var1}), (\ref{var2}) and (\ref{muvar})]. (b) The same for a  disk-shaped gas with 
$\nu =1, \lambda =10$. (c) The same for the trapped  Fermi gas of 1000 $^{161}$Dy atoms  
 versus the   parameter  $\lambda$ with $\nu=1$  from a numerical solution of   
    (\ref{3d1}) and variational approximation. 
%In all cases the oscillator length is $l_0= 1$
%$\mu$m.
}

\label{fig2}
\end{figure}

%For extreme trapping condition, when the system is of cigar or disk shapes, 
%we also solve the reduced 1D and 2D equations (\ref{1D}) and (\ref{2D}),
%respectively. % using the same procedure as above, e.g., the dipolar term %is 
%always treated in Fourier space by  FFT routines in reduced 1D and 2D equations. 
%The densities so obtained from the reduced 1D and 2D equations are compared %with 
%the solutions of the full 3D equation. %However, unlike in 3D, the Fourier 
%transformation of the 1D dipolar potential involving the integral $s_{1D}$ %of
%  (\ref{1Dvdd}) is evaluated numerically by the Gauss-Legendre quadrature 
%rule. The same transformation of the 2D dipolar potential involving 
%the function $h_{2D}$ of   (\ref{2Dvdd})
%is obtained essentially analytically using a Fortran library  
%subroutine for the error function erf. 

The anisotropic dipolar interaction is partially attractive in certain angles and repulsive in others and contributes very little  in a spherically 
symmetric trap. The dipolar interaction contributes attractively in the cigar-shaped configuration along the polarization $z$ direction and repulsively in the disk-shaped configuration perpendicular to the polarization $z$ direction. 
Hence we will mostly consider the degenerate dipolar Fermi gas in cigar and 
disk shapes. %By construction, in the linear limit $N\to 0$, the solutions of the reduced 1D and 2D equations (\ref{1D}) and (\ref{2D}) for cigar and disk shapes should yield results in agreement with the  3D model (\ref{3d1}). 
%It is interesting to explore how well the reduced 1D and 2D equations  perform for larger nonlinearities. 
%Compared to the bosonic GP equation, 
The 3D model (\ref{3d1}) is very strongly nonlinear with nonlinearity ${\cal N}=(6\pi^2 N)^{2/3}/2$, leading to large nonlinearities of ${\cal N}=760$ and 3526 for 
$N=1000$ and 10000, respectively.

First, we compare the results of the reduced 2D
density $n(x,y)$ of a disk-shaped degenerate Fermi gas of 
$^{161}$Dy atoms  with $\lambda=10, \nu=1$  as obtained from the 3D and 2D models (\ref{3d1}) and (\ref{2D}), respectively, and from the variational 
approximation to the 3D model. The reduced 2D density $n(x,y)$ is defined by $n(\rho)\equiv n(x,y)=\int_{-\infty}^{\infty} dz n(x,y,z)$. In  figure \ref{fig1} (a) we plot the reduced 2D density along $x$ axis $n(x,y=0)$ for different  $N$. Next, we compare the results of reduced 1D
density $n(z)$ of a cigar-shaped degenerate Fermi gas of 
$^{161}$Dy atoms  with $\lambda=0.1, \nu=1$  as obtained from the 3D and 1D models (\ref{3d1}) and (\ref{1D}), respectively, and from the variational 
approximation to the 3D model. The reduced 1D density $n(z)$ is defined by $n(z)=\int_{-\infty}^{\infty} dx \int_{-\infty}^{\infty}dy  n(x,y,z)$. In  figure \ref{fig1} (b) we plot the reduced 1D density along $z$ axis $n(z)$ for different  $N$. From  figure \ref{fig1} we find that in both cigar and disk shapes the 1D and 2D models perform fairly well, even for large nonlinearities, when compared with the results of the full 3D model. 
To see if the condition (\ref{con}) is satisfied by the shapes in  figure \ref{fig1}, we can use the TF estimate for $k_F$ of a trapped degenerate Fermi gas
\cite{frmp}: $k_F \approx (48N)^{1/6}/l_0$. For $100<N<10000$, as in  figure \ref{fig1},  the de Broglie wave length at the Fermi surface is 
$2\pi/k_F \approx l_0$ . The shapes and sizes in  figure \ref{fig1} are much
larger than this value, consistent with the condition (\ref{con}).
%, thus  justifying the applicability of the hydrodynamic equations. 

%It is appropriate to compare the variational approximations 
%(\ref{var1}), (\ref{var2}), and (\ref{muvar}) for the rms sizes and 
%chemical potential with those obtained from the solution of the full 3D 
%model (\ref{3d1}) with widely varying parameters. First, we 

%\begin{widetext}

%  \end{widetext}

Next we consider a 
cigar-shaped degenerate Fermi system of $^{161}$Dy atoms with trap 
parameters $\nu=1, \lambda =0.3$. In  figure \ref{fig2} (a) we plot the 
numerical and variational results for the chemical potential $\mu$ and 
rms sizes $\langle x \rangle $ and $\langle z \rangle $ of this system 
for  $10< N <10000$.  In  figure \ref{fig2} (b) 
we plot the same for a disk-shaped 
degenerate Fermi system with trap 
parameters $\nu=1, \lambda =10$. Finally, in  figure \ref{fig2} (c) we plot 
the same quantities for $N=1000$ versus the trap anisotropy $\lambda$ 
for $\nu=1$. In all cases %$-$ with widely varying trap anisotropy 
the variational results are in good 
agreement with the  3D model. {  For medium to 
small number ($N<10000$)
 of trapped $^{161}$Dy
atoms as considered here and also of experimental interest, 
the effect of the dipolar term in Eq. (\ref{3d1}) (the last term in this equation)
is small compared to the Fermi energy term (the last but one term there). Hence 
the effect of the dipolar term in figures \ref{fig2} is small. The plots in these 
figures only change by less than about two to four percent if we set $a_{dd}=0$. 
}
 
\subsection{Anisotropic sound propagation in uniform dipolar Fermi gas}

\begin{figure}[!b]
\begin{center}
\includegraphics[width=.7\linewidth]{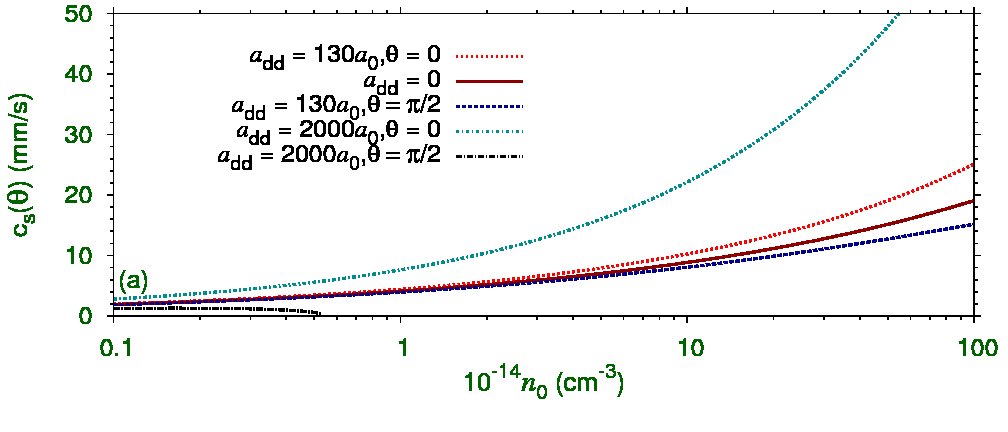}
\includegraphics[width=.7\linewidth]{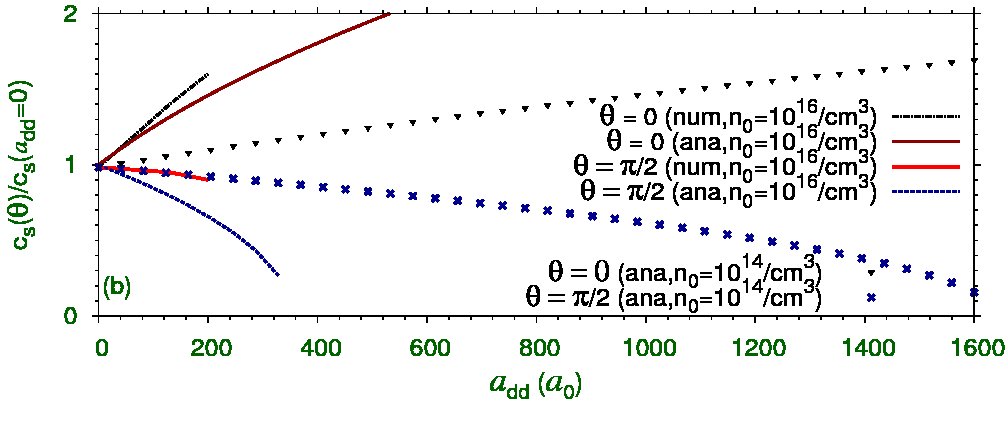}
\includegraphics[width=.7\linewidth]{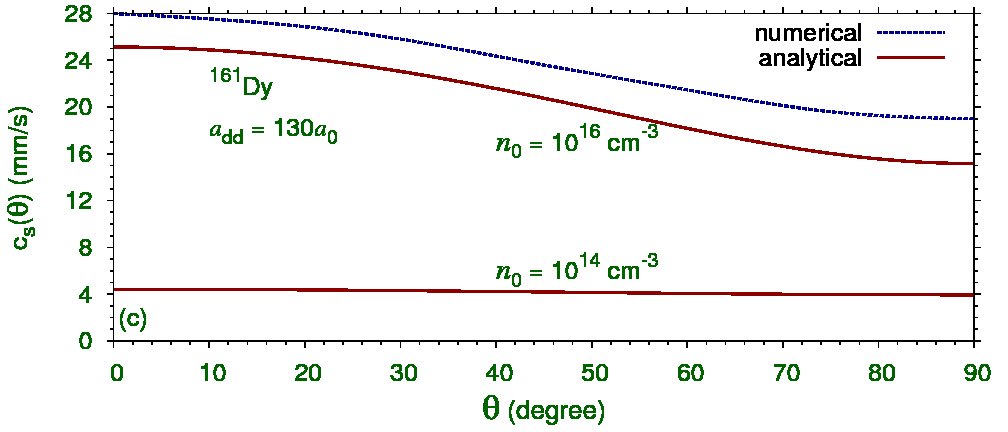}
\end{center}

\caption{ (a)  The analytical sound velocity $c_s(\theta)$
given by   (\ref{vel}) in axial $z$ ($\theta=0$) and radial $\rho$ ($\theta=\pi/2$) directions
versus density $n_0$ for $a_{dd}=130 a_0$ and 2000$a_0$ 
as well as the velocity for $a_{dd}=0$. 
 (b) The analytical (ana) sound velocity $c_s(\theta)$ in axial $z$ ($\theta=0$) and radial $\rho$ ($\theta=\pi/2$) directions versus $a_{dd}$
compared with the results of numerical simulation (num). 
(c) The analytical sound velocity $c_s(\theta)$ versus polar angle $\theta$ 
compared with the results of numerical simulation. 
}

\label{fig3}
\end{figure}

The hydrodynamic analytical result for sound velocity (\ref{vel}) in a 
uniform dipolar Fermi gas shows a clear anisotropy through the angle 
$\theta$ between the propagation direction $\bf r$ and the polarization 
direction $z$. In this expression for sound velocity there are two 
competing terms under the square root. The first term $v_F^2 /3$ 
involving the Fermi velocity is isotropic and proportional to 
$n_0^{2/3}$ whereas the second term proportional to dipolar interaction 
is anisotropic and proportional to density $n_0$. The anisotropy in 
sound velocity $c_s(\theta)$ will manifest strongly for large {
strength 
$a_{dd}$ and for large }
$n_0$ as 
the anisotropic dipolar term in   (\ref{vel}) will increase more 
rapidly with $n_0$ than the isotropic term. In  figure \ref{fig3} (a) we 
plot the sound velocity $c_s(\theta)$ for angles $\theta=0$ and $\pi$ as 
a function of density $n_0$ {for a degenerate Fermi gas with $a_{dd}=130a_0$
($^{161}$Dy atom)  and $2000a_0$ (polar $^{40}$K-$^{87}$Rb molecules in the singlet 
rovibrational ground state \cite{pfau})
 using the analytical result (\ref{vel}).} 
The sound velocity for 
nondipolar atoms ($a_{dd}=0$) is also plotted.  From this plot we find 
that the anisotropy in sound propagation as measured by the difference 
$[c_s(0)-c_s(\pi/2)]$ is sizable only for a relatively large density {$n_0=10^{15}$ cm$^{-3}$ for $^{161}$Dy atoms. However, for polar $^{40}$K-$^{87}$Rb molecules
the anisotropy is appreciable for a relatively low density of $n_0=10^{13}$ cm$^{-3}$.
From figure \ref{fig3} (a) we see that for $^{40}$K-$^{87}$Rb molecules, the sound velocity is imaginary 
for $\theta=0$ for a density of about $n_0=5\times10^{13}$ cm$^{-3}$.}
% In case of a dipolar BEC, the anisotropic sound propagation 
%can be prominent for a moderate density: $n_0= 10^{14}$ cm$^{-3}$ \cite{17}.

To study sound propagation,  the present stationary (static) 3D model (\ref{3d1}) is generalized to include time variation by replacing $\mu$ by the usual time derivative $i\hbar \partial/\partial t$.   Consequently, the infinite dipolar Fermi gas satisfies the following Galilei-invariant equation
\cite{as}
\begin{eqnarray}  \label{3d2} 
&& i\hbar \frac{\partial}{\partial t} \sqrt{n_0{(\bf r)}}
 =  \biggr[ -\frac{\hbar^2 \nabla^2 }{2(2m)} +2V({\bf r})
+2
\frac{\hbar^2}{2m}[6\pi^2  n_0({\bf r})]^{2/3}\nonumber \\
&&+ 12a_{dd}\frac{\hbar^2}{2m}\int \frac{1-3\cos^2 	\theta}{R^3}n_0({\bf r'})d{\bf r'}
\biggr] \sqrt{n_0({\bf r})}, 
\end{eqnarray} 
where density  $n_0({\bf r})=Nn({\bf r})$ is not normalizable for infinite hydrodynamics. 
Equation (\ref{3d2}) is consistent \cite{as}
with the time-dependent hydrodynamic 
equation (\ref{den})  after  the inclusion 
of the gradient correction term  \cite{grad}.

\begin{figure}[!b]
\begin{center}
\includegraphics[width=\linewidth]{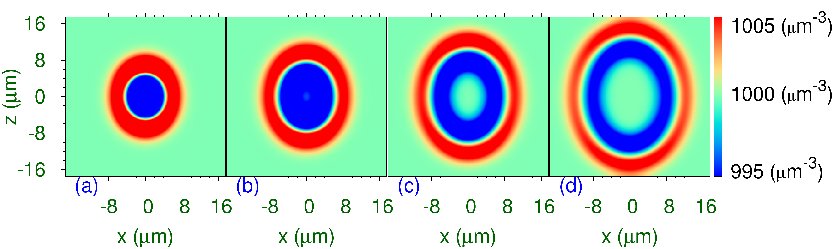}
\includegraphics[width=\linewidth]{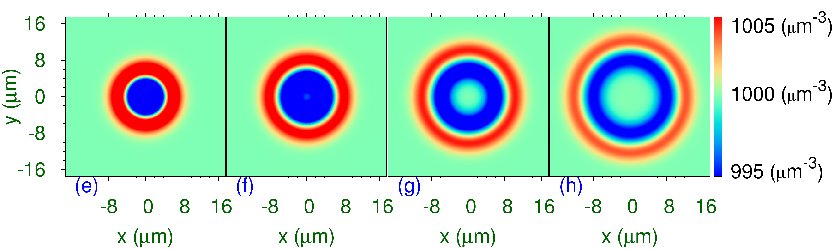}
\end{center}

\caption{ The contour plot of 3D density in $x-z$ plane 
$n_0(x,y=0,z)$ during sound propagation on a background density 
$n_0= 10^3$ $\mu$m$^{-3}$ of a degenerate dipolar $^{161}$Dy gas with $a_{dd}
=130a_0$ at times 
$t=$
(a) $0.15t_0$, (b) $0.2t_0$, (c) $0.25t_0$, and (d) $0.3t_0$ ($t_0=0.005$ s).
The contour plot of 3D density in $x-y$ plane 
   $n_0(x,y,z=0)$  during the same sound propagation at times 
(e) $0.15t_0$, (f) $0.2t_0$, (g) $0.25t_0$, and (h) $0.3t_0$. 
}

\label{fig4}
\end{figure}

{ The anisotropy of the dipolar interaction would be prominent at low to medium density for large 
dipolar interaction.  To illustrate the anisotropic sound propagation we will consider two examples: 
$^{161}$Dy atoms at a medium density of 10$^{15}$ cm$^{-3}$ and the polar molecules 
$^{40}$K-$^{87}$Rb at the low density of 10$^{13}$ cm$^{-3}$.
First we consider 
the numerical simulation of sound propagation in the infinite   $^{161}$Dy gas 
($a_{dd}=130a_0$)
at a background density of
$n_0=10^3 $ $\mu$m$^{-3}=10^{15}$ cm$^{-3}$.  }
The numerical simulation is initiated with 
a 3D Gaussian pulse
at the center of the uniform 3D background
density given by $n_0({\bf r})
 = (10^3 + 10^2e^{-r^2/2w^2})$ $\mu$m$^{-3}$, $w=2 $ $\mu$m,
subject to a weak expulsive Gaussian potential
$V ({\bf r}) = 0.00001e^{-r^2/2w^2}$ $\mu$m$^{-2}$. 
With this initial condition   (\ref{3d2}) is solved by
 real-time propagation \cite{CPC} to study sound waves.
An ellipsoid-like sound wave front is found to emerge outwards upon time propagation. From a numerical study of this wave front the sound velocity 
in different directions is calculated.

Typical anisotropic sound propagation in a $^{161}$Dy degenerate gas
of density 10$^{15}$ cm$^{-3}$ is shown in  figure \ref{fig4}. 
The sound propagation is illustrated via contour plots of 3D density $n_0(x,y,0)$
in the $x-y$ plane 
 and of density $n_0(x,0,z)$ in the $x-z$ plane. 
In  figures \ref{fig4} (a), (b), (c), and (d), to illustrate 
the anisotropic sound propagation in the $x-z$ plane, 
we show the contour plot of  $n_0(x,0,z)$ at times $t= 0.15t_0, 0.2t_0,
0.25t_0$ and $0.3t_0$, respectively,  with 
$t_0=2ml_0^2/\hbar^2\approx 0.005$ s the time scale and 
$l_0=1$ $\mu$m the 
length scale used in the numerical solution of 
  (\ref{3d2}). However, the propagation in the radial $x-y$ plane 
is isotropic. This isotropic  sound propagation is shown in  figures 
\ref{fig4} (e), (f), (g), and (h) 
via the contour plot of  $n_0(x,y,0)$
at times $t= 0.15t_0, 0.2t_0,0.25t_0$ and $0.3t_0$, respectively. 
A clean wave front of high density is identified in these contour 
plots encompassing a region of low density $-$ a typical panorama in sound propagation. 
%For the background density of $n_0=10^4$ $\mu$m$^{-3}$, the de Broglie wave length is
%$2 \pi/k_F\approx 2 \pi/(6\pi^2n_0)^{1/2}\approx 0.1 $ $\mu$m is much smaller than the length scales of  figure \ref{fig4}, in agreement with condition (\ref{con}).

\begin{figure}[!t]
\begin{center}
\includegraphics[width=\linewidth]{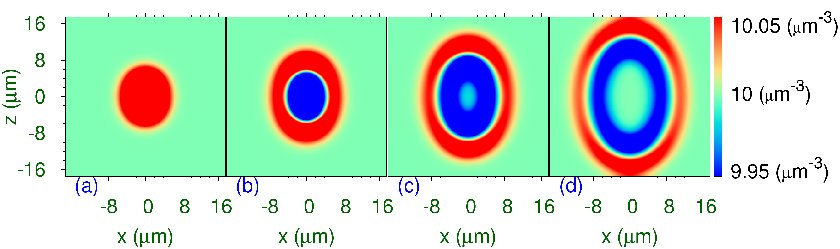}
\includegraphics[width=\linewidth]{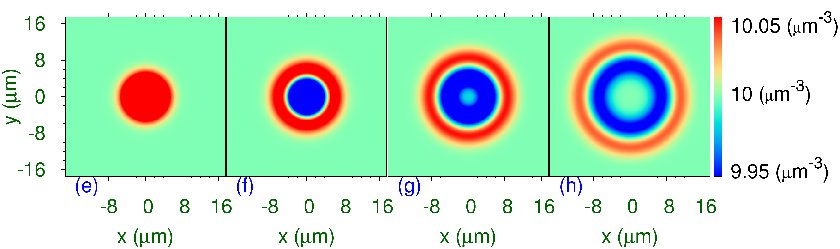}
\end{center}

{
\caption{  The contour plot of 3D density in $x-z$ plane 
$n_0(x,y=0,z)$ during sound propagation on a background density 
$n_0= 10$ $\mu$m$^{-3}$ of a degenerate dipolar molecular $^{40}$K-$^{87}$Rb gas with $a_{dd}
=1000a_0$ at times 
$t=$
(a) $0.2t_0$, (b) $0.5t_0$, (c) $0.8t_0$, and (d) $1.1t_0$ ($t_0=0.004$ s, $l_0=1$ $\mu$m).
The contour plot of 3D density in $x-y$ plane 
   $n_0(x,y,z=0)$  during the same sound propagation at times 
(e) $0.2t_0$, (f) $0.5t_0$, (g) $0.8t_0$, and (h) $1.1t_0$. }}

\label{fig5}
\end{figure}

The sound propagation
along $z$ direction  (polar angle $\theta=0$)
has a velocity ($c_s=v_F/\sqrt 3$) larger than that 
for a nondipolar system ($a_{dd}=0$)
of same density, whereas that in the $x-y$ plane  
(polar angle $\theta =\pi/2$)
has a velocity smaller than that for a nondipolar
system as can be seen from   (\ref{vel}). 
The analytical  sound velocity $c_s$
for nondipolar atoms  of density $10^{15}$ cm$^{-3}$ and atomic mass 161
is 8.88 mm/s compared to the numerically obtained  velocity of 8.5 mm/s. 
For $^{161}$Dy atoms of density $n_0=10^{15}$ cm$^{-3}$
with $a_{dd} = 130a_0$, 
the analytical radial sound velocity
is $c_s(\pi/2) = 8.1$ mm/s (numerical 7.0 mm/s), and
the analytical axial  sound velocity is $c_s(0) = 10.3$ mm/s
 (numerical 10.0 mm/s). In  figure \ref{fig3} (b), we present the variation 
of axial ($\theta =0$) and radial ($\theta=\pi/2$) sound velocities 
versus $a_{dd}$ as obtained from numerical simulation and analytical 
consideration (\ref{vel}). In  figure \ref{fig3} (c), the analytical result 
for velocity versus the polar angle $\theta$ is compared with the 
numerical result. The agreement between 
the analytical result 
(\ref{vel}) for sound velocity and the result of numerical simulation is 
satisfactory considering the very large nonlinearities present in the 
system due to large density ($n_0$) and large dipolar interaction.

{
To have a larger anisotropy in sound propagation than in a gas of $^{161}$Dy atoms at a lower density 
a stronger dipolar interaction, as in polar molecules, is needed. A convenient polar fermionic molecule 
molecule  $^{40}$K-$^{87}$Rb has  
a permanent electric dipole moment 0.052 Debye ($a_{dd}\approx 20a_0$) for the triplet rovibrational ground state and 0.566 Debye ($a_{dd}\approx 2000a_0$) for the singlet rovibrational ground state \cite{pfau}.  For illustration, we consider a uniform gas of $^{40}$K-$^{87}$Rb molecules of density $n_0=10^{13}$  cm$^{-3}$ with   $a_{dd} = 1000a_0$. 
The numerical simulation is initiated with 
a 3D Gaussian pulse
at the center of the uniform 3D background
density given by $n_0({\bf r})
 = (10 + e^{-r^2/2w^2})$ $\mu$m$^{-3}$, $w=2 $ $\mu$m,
subject to a weak expulsive Gaussian potential
$V ({\bf r}) = 0.00001e^{-r^2/2w^2}$ $\mu$m$^{-2}$. 
With this initial condition   (\ref{3d2}) is solved by
 real-time propagation \cite{CPC} to study sound waves.
Typical anisotropic sound propagation in a $^{40}$K-$^{87}$Rb degenerate gas
of density 10$^{13}$ cm$^{-3}$ is shown in  figure 5. 
Because of the stronger dipolar interaction, a larger anisotropy has appeared
at a lower density in 
figure 5 compared to figure \ref{fig4}. For $^{40}$K-$^{87}$Rb molecules 
of density $n_0=10^{13}$ cm$^{-3}$
with $a_{dd} = 1000a_0$, 
the analytical radial sound velocity
is $c_s(\pi/2) = 2.05$ mm/s (numerical 2.4 mm/s), and
the analytical axial  sound velocity is $c_s(0) = 3.03$ mm/s
 (numerical 3.4 mm/s). 
 }

\label{IIIB}

\section{Summary}

We developed a 3D theoretical formulation for describing certain macroscopic observables of a degenerate dipolar Fermi gas appropriate for studying stationary properties, such as, shape, size, energy, chemical potential etc. of a trapped system. The effect of dipolar interaction is negligible in the spherically-symmetric configuration and
the dipolar interaction manifests strongly in the asymmetric cigar and disk shapes. Simple reduced equations in 1D and 2D suitable for studying the trapped degenerate dipolar Fermi gas in cigar and disk shapes, respectively,
are derived. Also, a Gaussian variational approximation for studying these macroscopic properties is derived.  We apply the present formalism to study the stationary properties of a trapped degenerate Fermi gas of $^{161}$Dy atoms. The stationary properties of the 3D model under appropriate conditions are found to be in satisfactory agreement with those from the reduced 1D and 2D models as well as with variational approximation.

Using the present 3D model we also obtained analytical results for 
anisotropic sound velocity 
as a consequence of the anisotropic dipolar interaction
in a dipolar Fermi gas in agreement with 
numerical simulation of the 3D   (\ref{3d2}).   
The sound velocity is larger along the polarization direction than 
in the transverse plane.

\ack

%\acknowledgments
We thank FAPESP  and  CNPq (Brazil)  for  support.
We thank
Luca Salasnich for useful comments.

\end{document}